# L3CCD results in pure photon counting mode

Olivier Daigle[*a,b], Jean-Luc Gach[b], Christian Guillaume[c], Claude Carignan[a], Philippe Balard[b], Olivier Boissin[b]

[a]Laboratoire d'Astrophysique Expérimentale & Département de physique, Université de Montréal, C.P. 6128, Succ. Centre-ville, Montréal, Québec, Canada H3C 3J7;

[b]Observatoire de Marseille, 2, Place Le Verrier, 13248 Marseille cedex 4, France;

[c]Observatoire de Haute-Provence, 04870 St-Michel l'observatoire, France.

**ABSTRACT**

Theoretically, L3CCDs are perfect photon counting devices promising high quantum efficiency (~90%) and sub-electron readout noise ($\sigma < 0.1$ e[-]). We discuss how a back-thinned 512x512 frame-transfer L3CCD (CCD97) camera operating in pure photon counting mode would behave based on experimental data. The chip is operated at high electromultiplication gain, high analogic gain and high frame rate. Its performance is compared with a modern photon counting camera (GaAs photocathode, QE ~28%) to see if L3CCD technology, in its current state, could supersede photocathode-based devices.

## 1. INTRODUCTION

Photon counting (PC) efficiency has increased since the first photon counting devices (Boksenberg 1977, Blazit et al. 1977). Now, third generation image intensifiers with high quantum efficiency (QE) photocathodes (of the order of 30%) are used with CCDs in modern intensified photon counting systems (IPCS) (Gach et al. 2002, Hernandez et al. 2003). But, the efficiency of this kind of camera is limited by the capability of the tube to transform an incoming photon into an electron and amplifying it so that the output signal is largely over the CCD readout noise. Improvements in CCD quantum efficiency, which now reaches 90% at some wavelengths, and readout noise (RON), as low as 3e[-], have no effect on the total counting efficiency of these cameras. RON of 3e[-] is still too high for an application such as photon counting for extreme faint fluxes.

It is now possible to amplify the pixel signal into the CCD before it reaches the output amplifier and before it is affected by its noise (eg, Jerram et al.). Gain is created by adding a special shift register excited with a high voltage phase that makes avalanche multiplication probable at every shift. The probability of amplification being small (p<2% and variable depending on the voltage applied and temperature), the amount of elements (N>500) makes the total gain large (G=$p^N$ being in the range of thousands). Interestingly, the *effective* RON may be seen as the real readout noise divided by the amplification factor that has been applied to the pixel's charge. Hence,

$$\sigma_{eff} = \frac{\sigma_{real}}{G} \quad , \tag{1}$$

where *G* is the gain that is applied to the charge before it reaches the output amplifier.

If one uses such a CCD in high gain mode to get a $\sigma_{eff}$ in the range of 0.2, one could think that by applying a 5$\sigma$ detection threshold this would make photon counting possible. But, as this gain is statistical, there is only a *probability* of amplification at every stage of the shift register. As a result, the exact gain that has been applied to a pixel's signal when it reaches the output amplifier is impossible to tell. This means that at high gains, needed for photon counting, the pixel becomes nonlinear. It is impossible to accurately determine from the output signal value how many events occurred.

In multiplex observations, a conventional CCD signal to noise ratio (SNR) may be expressed as follows,

---

* odaigle@astro.umontreal.ca; phone +1 514 343-6111 #1663; fax +1 514 343-2071

$$SNR = \frac{S}{\sqrt{S + T + n\sigma^2}} \quad , \tag{2}$$

where $S$ is the amount of electrons collected per pixel during the all the integration periods, $T$ the thermal noise in electrons•pixel$^{-1}$, $n$ the number of readouts and $\sigma$ the readout noise of the CCD in electrons. This SNR formula needs some changes to reflect the SNR of an Electron Multiplying CCD (EMCCD). First, the readout noise is as shown by equation (1). Second, the stochastic multiplication add noise by convoluting the pixel's signal and is reflected by a noise factor given by (Stanford and Hadwen, 2002)

$$F = \sqrt{\frac{2(G-1)}{G^{\frac{N+1}{N}}} + \frac{1}{G}} \quad . \tag{3}$$

When $G$ is large, $F^2 = 2$. With these two modifications in mind, one can thus rewrite the SNR equation for a theoretical EMCCD as

$$SNR = \frac{S}{\sqrt{F^2 S + F^2 T + n\frac{\sigma_{real}^2}{G^2}}} \quad . \tag{4}$$

## 2. L3CCD OPERATION FOR PHOTON COUNTING

When the L3CCD is operated at high gain (G>100), its noise factor, as shown by equation (3), reaches a value of $2^{\frac{1}{2}}$ and it has the same effect as halving the QE. But, this noise factor can be neglected if the output of the CCD is considered binary, that is, if the output is above a given level the pixel will be considered as having undergone a single event during the integration time. On the contrary, a lower output level will be considered as having not undergone an event. Using this simple thresholding scheme, one can recover the full silicon QE of the L3CCD, making it a potentially perfect detector.

Care must be taken when using this thresholding method. The threshold level must not be set too high, as shown in Figure 1, since it would result in losing some events and decreasing the resulting QE of the system. Also, if more than one event occurs on a pixel during an integration, the exceeding events will be lost, resulting in a linearity loss. The frame rate must thus be set according to the expected flux to avoid losing events as much as possible. This linearity loss may be compensated out as the mean number of missed photons can be evaluated (Gach et al. 2002). Since photon emission can be assumed as being a Poissonian process, the following equation

$$g = \frac{1 - e^{-\alpha}}{\alpha} \quad , \tag{5}$$

gives $g$, the proportion of counted photons as a function of $\alpha$, the mean number of photons expected during the integration period. As a result, this correction adds a nonlinear noise that scales as

$$N = \sqrt{\frac{\alpha}{g}} \quad . \tag{6}$$

To avoid being affected too much by this noise, one must choose a frame rate that will give a maximum mean signal of $10^{-1}$ photon pixel$^{-1}$ frame$^{-1}$, as shown by Figure 2. With that in mind, the threshold level must be chosen according to the readout noise of the CCD. The threshold must be high enough to be out of the RON of the CCD and avoid creating false events and low enough to avoid losing real events as previously stated.

Since the SNR is dependent of the flux, frame rate and QE and that the QE is dependent on the threshold, the ideal threshold level will depend on the flux per pixel per integration. To properly determine the threshold, it is interesting to plot the effective SNR of an observation according to the threshold used for various light levels.

Figure 3 shows that a threshold level of 5.5$\sigma$ is where the resulting SNR will be at its best to cover a wide dynamic range. This figure plots curves for a system running at 30fps with a RON of 30e$^-$, but the best threshold is independent of the RON as long as it is expressed in sigma. The relative SNR, where all low fluxes meet, is always situated at ~5.5$\sigma$ (5.42 precisely, determined empirically). If the expected flux is known and its dynamic range is small, one could choose a threshold lower than 5.5σ to get a small gain in SN.

Using a threshold of 5.5σ, one can now determine the gain at which the L3CCD must be operated according to its system RON to lower the loss in effective QE, as shown in Figure 4. A good approximation would be that the gain must be at least 100 times greater than the RON of the system to avoid losing too much QE. Since gains of the order of 5000 requires very low temperatures (<173K) and maximum high voltage clock amplitude (46V), this implies that an L3CCD operating in photon counting mode should have less than 50e- of RON.

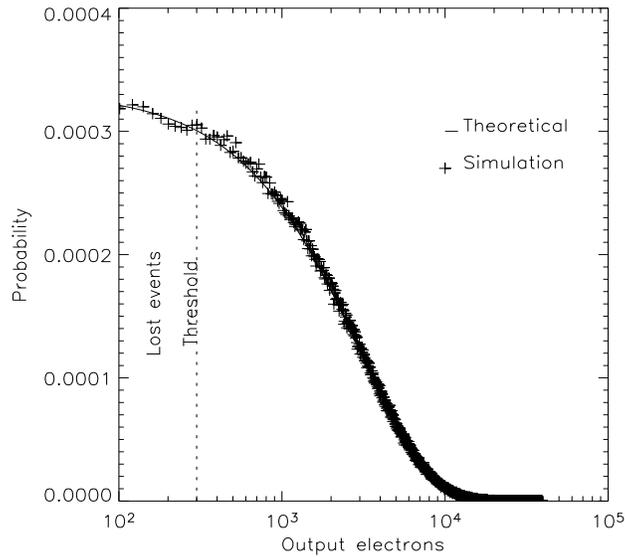

*Figure 1 : Output probability for a single electron passing through a 536 elements register each having a multiplication probability of 1.5%, resulting in a mean gain of ~3000.*

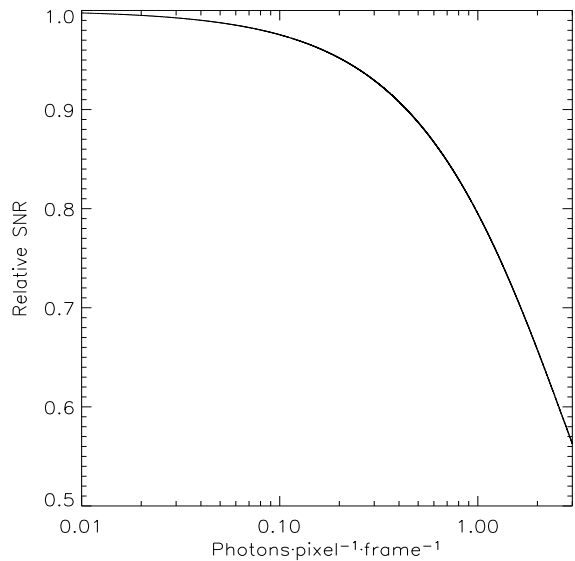

*Figure 2 : Simulation showing the decrease in SNR due to nonlinearities as the flux is increased.*

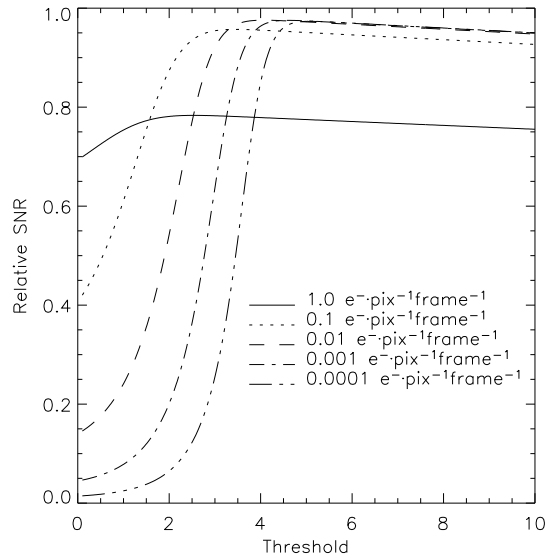

*Figure 3 : Relative SNR versus threshold level used (in sigma) assuming a RON of 30e⁻ for different fluxes. SNR is normalized according to a perfect system that would not need any thresholding and would be perfectly linear.*

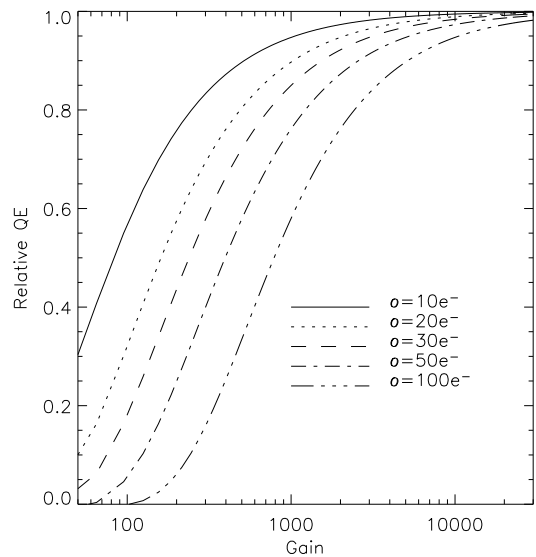

*Figure 4 : Relative QE versus gain for different RON and using a threshold of 5.5σ. QE is normalized according to a system without any RON.*

## 3. REAL LIFE L3CCD NOISE

### 3.1. Clock induced charges

When it comes to photon counting with L3CCDs, clock induced charges (CIC), a noise source that could be neglected on conventional CCDs, arises (Jerram et al.). CIC appear during the transfer of charges through the device and are usually low enough to be buried in the readout noise of conventional CCDs. But, when the L3CCD is operated at high gain, the induced charges will appear at the output amplifier as a normal event, thus creating a false event impossible to discriminate from a real event. These false events will thus add noise to the image. Even if this noise is generated in both serial and parallel registers, it seems that the vertical component dominates over the serial one under most operating conditions (E2V Technologies, 2004).

The total amount of charges induced during the vertical shift process will not be equal for all lines since they will not undergo the same amount of transfers. This will create a gradient of CIC across the CCD and the last readout line will have roughly twice as much CIC as the first one (for frame-transfer CCDs). For the sake of simplicity, one may consider the mean amount of CIC that will appear on pixels across the CCD and rewrite equation (4) as follows

$$SNR = \frac{S}{\sqrt{F^2 S + F^2 T + nF^2 C + \frac{n\sigma_{real}^2}{G^2}}}, \quad (7)$$

where $C$ is the mean CIC per pixel per readout in electrons.

Since the amount of CIC varies according to the clock levels, rise time, fall time and operation mode of the CCD (inverted, non-inverted), it is interesting to compare its effect on SNR according to light level. Doing so will enable one to determine if the expected amount of CIC will ruin the SNR of a given application. Figure 5 shows clearly that for typical photon counting applications where high temporal resolution is needed, the CIC must be kept very low, below $10^{-4}$ e$^-$ pixel$^{-1}$ frame$^{-1}$.

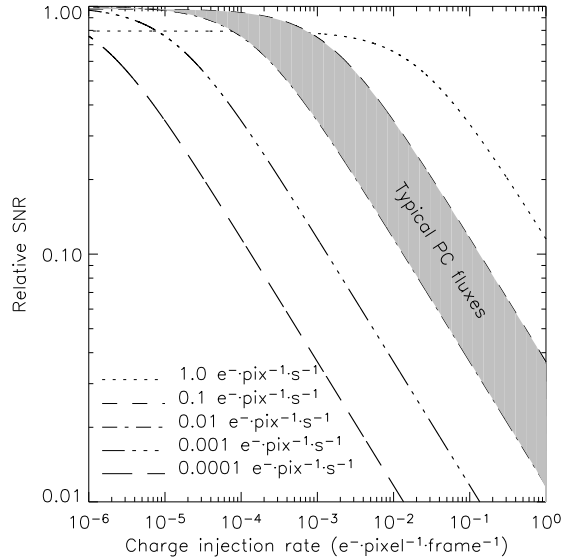

*Figure 5 : Relative SNR according to CIC level at 30 fps in photon counting mode. SNR is normalized according to a system that would not have any CIC, would not be affected by a threshold and would be perfectly linear.*

If the temporal resolution is less important, the L3CCD could be operated at a lower frame rate. Since typical photon counting applications involve fluxes in the range of $10^{-2}$ to $10^{-1}$ photon pixel$^{-1}$ frame$^{-1}$, a frame rate of 1 fps could be enough. Dropping the frame rate from 30 fps to 1fps involves having 30 times less CIC, which will greatly raise the SNR of the observation. Figure 6 shows that in that case, the CIC level must be kept below $3\times10^{-3}$ e$^-$ pixel$^{-1}$ frame$^{-1}$. The figure also shows how the nonlinearity of the lower frame rate affects the resulting SNR. If a flux larger than $10^{-1}$ photon pixel$^{-1}$ frame$^{-1}$ is expected, the frame rate should absolutely be raised to recover SNR.

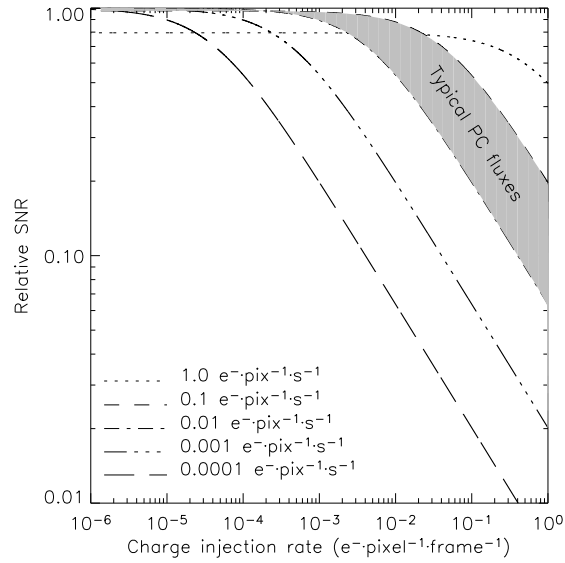

*Figure 6 : Relative SNR according to CIC level at 1 fps in photon counting mode. SNR is normalized according to a system that would not have any CIC, would not be affected by a threshold and would be perfectly linear.*

### 3.2. Dark signal

Dark signal of an L3CCD behaves exactly as the one of a classical CCD. When operated in inverted mode, the dark signal is only composed of the signal generated in the bulk. But, as charge injection is diminished by operating the CCD in non-inverted mode (E2V Technologies, 2004), one may find interesting to operate the CCD in that mode at the price of a thermally generated signal at the silicon surface that will add itself to the signal generated in the bulk. That is,

$$S_D = S_S + S_B \quad , \tag{8}$$

where $S_D$ is the total dark signal, $S_S$ is the surface dark signal and $S_B$ is the bulk dark signal. Using data from E2V Technologies, it is possible to approximate that when operated in non-inverted mode, the surface dark signal is 2 orders of magnitude greater than the bulk dark signal. Since a bulk dark signal of $10^{-3}$ e$^-$ pixel$^{-1}$ second$^{-1}$ is typical on CCDs when deep cooling (<173K) is applied, one may postulate that the surface dark signal will be of the order of $10^{-1}$ e$^-$ pixel$^{-1}$ second$^{-1}$ in that case. When operating in inverted mode, $S_S$ simply vanishes.

Taking into account that typical CIC for a CCD97 operating in non-inverted mode is $3\times10^{-6}$ e$^-$ pixel$^{-1}$ transfer$^{-1}$ and that every line undergo an average of 804 vertical transfers (536 for the frame transfer and 536/2 for the readout), the resulting mean CIC is of the order of $2.4\times10^{-3}$ e$^-$ pixel$^{-1}$ frame$^{-1}$ in that mode. When operating in inverted mode, CIC would be more typically of the order of $8\times10^{-2}$ e$^-$ pixel$^{-1}$ frame$^{-1}$. Since dark signal is time dependent and CIC is frame rate dependent, one may show that for high frame rates it is best to operate the L3CCD in non-inverted mode since CIC dominates. For low frame rates it is best to operate the L3CCD in inverted mode since dark current dominates. Figure 7 shows that using the data provided, dark signal dominates CIC when operated in non-inverted mode for frame rates lower than 1.3fps. So, one must choose to operate the L3CCD in inverted mode for frame rates lower than that threshold and in non-inverted mode of frame rates higher. In any case, deep cooling is required in photon counting mode in order to keep the overall dark signal as low as possible.

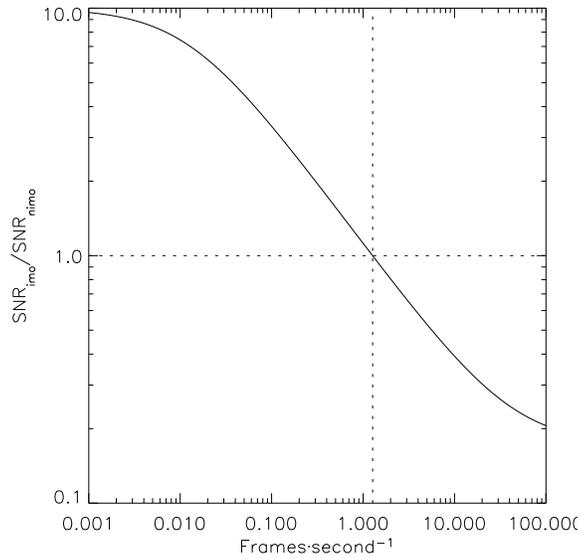

*Figure 7 : Relative SNR between inverted and non-inverted mode operation in function of the frame rate. For IMO, CIC is $8\times10^{-2}$ $e^-$ $pixel^{-1}$ $frame^{-1}$, dark signal is $10^{-3}\times e^-$ $pixel^{-1}$ $second^{-1}$. For NIMO, CIC is $2.4\times10^{-3}$ $e^-$ $pixel^{-1}$ $frame^{-1}$, dark signal is $10^{-1}$ $e^-$ $pixel^{-1}$ $second^{-1}$. Deep cooling must be applied.*

## 4. DISCUSSION

It has been shown that L3CCD performance in photon counting mode is affected by four general terms (QE loss, nonlinear noise, CIC and dark signal) which themselves are function of different factors (RON, threshold, gain, frame rate, flux, temperature and operation mode). This is summarized in the following table :

| is function of | RON | Threshold | Gain | Frame rate | Flux | T° | Operation mode |
|---|---|---|---|---|---|---|---|
| QE loss | x | x | x | | | | |
| Nonlinearity | | | | x | x | | |
| CIC | | | | x | | | x |
| Dark signal | | | | | | x | x |

*Table 1 : Factors that affects L3CCD performance and their impact on general terms.*

CIC and dark signal are closely related since the operating mode of the L3CCD will affect them both. By using Figure 7, one may now determine the best operation mode for the frame rate that is needed. Now, considering that the noise of a perfect photon counting system would be only shot noise, its SNR would be

$$SNR_{perfect} = \frac{S}{\sqrt{S}} = \sqrt{S} \quad . \tag{9}$$

With that in mind it is possible to compare various L3CCD configurations by plotting them against a perfect photon counting system in order to determine which is the best for any given application. Figure 8 compares three different L3CCD configurations together with a classical back-thinned CCD and an GaAs IPCS (Gach et al. 2002, Hernandez et al. 2003).

By taking into account all factors that affects the SNR apart from CIC and dark signal, it is possible to show that the final SNR of a perfect L3CCD having a RON of 30$e^-$ operating at a gain of 3000 (loss of 5.8% in QE), having a threshold of 5.5σ (loss of 2.8% in SNR) and that is readout at a rate at least 10 times higher than the maximum expected light flux (loss of 2.5% in SNR) would not differ more than 8.4% from a perfect photon counting system for the same QE. But, by looking at Figure 8, one sees how badly the sum of CIC and dark signal affects the L3CCD. In fact, even if the QE is three times larger, it can hardly compete with a GaAs IPCS for typical real-time photon counting applications, where expected fluxes are lower than $10^{-1}$ photon pixel$^{-1}$ second$^{-1}$. For fluxes higher than $10^{-1}$ and lower than 10 photons pixel$^{-1}$ second$^{-1}$, the L3CCD operating in non-inverted mode at 30fps is clearly an advantage because of its high QE and because the CIC and dark signal become negligible as compared to the incoming flux. The figure also shows the effect of the nonlinear noise of the L3CCDs in photon counting mode when the flux goes higher than the frame rate.

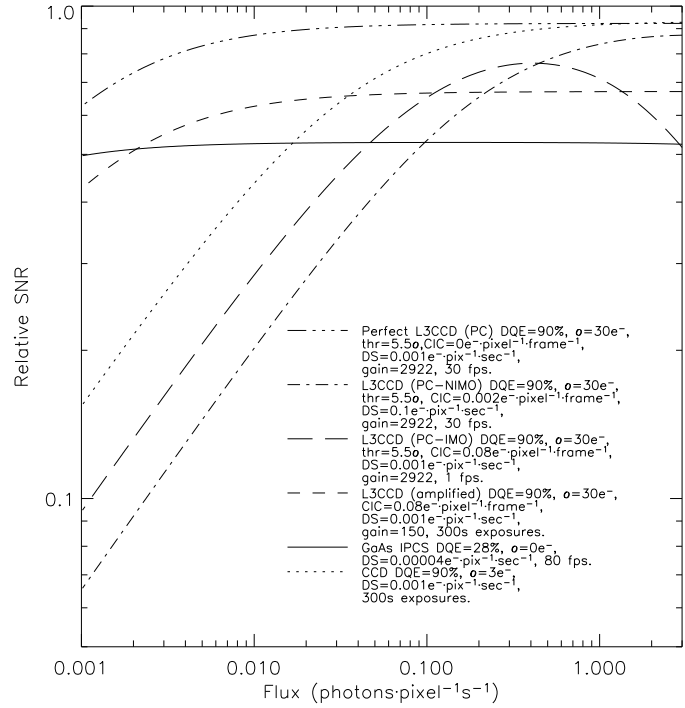

*Figure 8 : SNR comparison of different detectors under different light fluxes. SNR is normalized to a perfect photon counting system (100% QE and no readout noise).*

Regarding applications where the real-time factor is not needed, the L3CCD operating in amplified mode and readout every 5 minutes is really an advantage as compared to classical back-thinned CCD for very low flux applications ($<3\times10^{-2}$ photon pix$^{-1}$ second$^{-1}$). For fluxes higher than that, the performance of a classical CCD is higher since it does not suffer from the noise factor of the L3CCD.

## 5. CONCLUSION

As long as CIC and dark signal will be as strong as they currently are, L3CCDs will not be able to compete with GaAs IPCS for real-time applications. One way of reducing CIC would be to have many outputs so that every line would not be shifted too many times to be readout. Of course, this makes it impossible to have a large frame transfer L3CCD as all the lines must be shifted many times to be brought into the storage section. This also makes it impossible to have small integration times that are of the order of the readout time since it would be hard to use a shutter, not to mention that the time spent with the shutter closed transposes in QE loss. For very small L3CCDs, or for rectangular L3CCDs, then it would be possible to have a storage section placed on both sides of the image section so that the amount of shifts to bring the signal into the storage section would be minimized.

From another point of view, since the GaAs QE wavelength range is limited (Figure 9), L3CCD is more sensitive at very blue and very red wavelengths. For real-time applications at fluxes higher than $10^{-2}$ photon pix$^{-1}$ second$^{-1}$, it is only when the L3CCD QE is 10 times higher than GaAs QE ($\lambda$<420nm and $\lambda$>900nm) that it would be more efficient to use an L3CCD than a GaAs IPCS for photon counting.

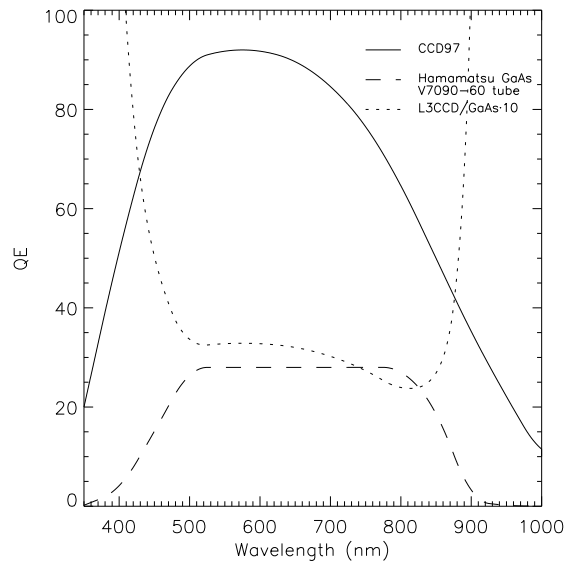

*Figure 9 : L3CCD QE compared to an Hamamatsu V7090-60 GaAs amplification tube. The dotted line shows the relative L3CCD/GaAs QE multiplied by 10 for the sake of clarity.*

## ACKNOWLEDGMENTS

We would like to thank the European Southern Observatory, the Observatoire du mont Mégantic, the Fondation Canadienne de l'Innovation, the Centre National de la Recherche Scientifique, the Université de Provence and the Université de Montréal for funding this study.